\IfFileExists{revtex4-2.cls}{%
  \documentclass[aps,prd,twocolumn,showpacs,floatfix,amsmath,amssymb,nofootinbib,superscriptaddress,,preprintnumbers]{revtex4-2}%
}{%
  \documentclass[aps,prd,twocolumn,showpacs,floatfix,amsmath,amssymb,nofootinbib,superscriptaddress,preprintnumbers]{revtex4-1}%
}

\usepackage{graphicx}
\usepackage{bm}
\usepackage{xcolor}
\usepackage[colorlinks=true,linkcolor=blue,citecolor=blue,urlcolor=blue]{hyperref}
\usepackage{fontawesome}

\graphicspath{{./}{figures_prl/}{figures/}}

\newcommand{\be}{\begin{equation}}
\newcommand{\ee}{\end{equation}}
\newcommand{\dd}{\mathrm{d}}
\newcommand{\UL}{\mathrm{UL}}
\newcommand{\Lensed}{\mathrm{L}}
\newcommand{\MLz}{M_{\mathrm{L}z}}
\newcommand{\msun}{M_\odot}

\begin{document}

\title{Searching for lensed gravitational waves with unlensed templates}

\author{Digvijay Wadekar}
 \email{jay.wadekar@utexas.edu}
\affiliation{\mbox{Center for Gravitational Physics, University of Texas at Austin, Austin, TX 78712, USA}}
\author{Barak Zackay}
\affiliation{\mbox{Department of Particle Physics \& Astrophysics, Weizmann Institute of Science, Rehovot 76100, Israel}}
\author{Mark Ho-Yeuk Cheung}
\affiliation{\mbox{School of Natural Sciences, Institute for Advanced Study, 1 Einstein Drive, Princeton, NJ 08540, USA}}
\author{Tousif Islam}
\affiliation{\mbox{Center for Gravitational Physics, University of Texas at Austin, Austin, TX 78712, USA}}
\author{Javier Roulet}
\affiliation{\mbox{Kavli Institute for Cosmological Physics, The University of Chicago, 5640 South Ellis Avenue, Chicago, Illinois 60637, USA}}
\affiliation{\mbox{School of Natural Sciences, Institute for Advanced Study, 1 Einstein Drive, Princeton, NJ 08540, USA}}
\author{Ajit Kumar Mehta}
\affiliation{\mbox{Chennai Mathematical Institute, Siruseri, Tamil Nadu 603103, India}}
\affiliation{\mbox{International Centre for Theoretical Sciences, Tata Institute of Fundamental Research, Bangalore 560089, India}}
\author{Tejaswi Venumadhav}
\affiliation{\mbox{Department of Physics, University of California at Santa Barbara, Santa Barbara, CA 93106, USA}}
\affiliation{\mbox{International Centre for Theoretical Sciences, Tata Institute of Fundamental Research, Bangalore 560089, India}}
\author{Jonathan Mushkin}
\affiliation{\mbox{Department of Particle Physics \& Astrophysics, Weizmann Institute of Science, Rehovot 76100, Israel}}
\author{Miguel Zumalac\'{a}rregui}
\affiliation{\mbox{Instituto de F\'{\i}sica Te\'{o}rica UAM/CSIC, C/ Nicol\'{a}s Cabrera 13--15, Universidad Aut\'{o}noma de Madrid,}\\\mbox{Cantoblanco, 28049 Madrid, Spain}}
\affiliation{\mbox{Max Planck Institute for Gravitational Physics (Albert Einstein Institute), Am M\"{u}hlenberg 1, D-14476 Potsdam, Germany}}
\author{Matias Zaldarriaga}
\affiliation{\mbox{School of Natural Sciences, Institute for Advanced Study, 1 Einstein Drive, Princeton, NJ 08540, USA}}

\date{\today}

\begin{abstract}

Gravitational lensing can split the signal of a binary merger into two images, which either overlap and interfere or arrive as separate events. We show that the matched-filter output of every two-image template can be straightforwardly constructed from the signal-to-noise ratio (SNR) time series that searches already compute with unlensed templates. This enables a marginalized detection statistic that increases search sensitivity without additional matched filtering and is more than \(150\) times faster than an explicit lensed template bank. For overlapping images in the geometric-optics regime, our two-image method recovers $\gtrsim97\%$ of the optimal SNR from point-mass and singular isothermal lenses, compared with as little as $67\%$ captured by unlensed templates, with substantial gains also for Chang--Refsdal and cored isothermal lenses. Overlapping lensed signals moreover fail the signal-consistency test that is used to down-rank noise transients. Our two-image method restores the consistency and increases the sensitive volume by roughly $2$--$4\times$ for the strongly down-ranked signals. For separated images, our method enables an efficient coherent search for pairs in which neither image is significant alone. Near fold caustics, this increases the sensitive volume by factors of \(1.8\)--\(2.6\) over requiring two independent detections and up to \(1.6\) over a targeted counterpart search. Our method is applicable to any matched-filtering search. \href{https://github.com/JayWadekar/lensed-gw-unlensed-templates}{\faGithub}

\end{abstract}

\preprint{IFT-UAM/CSIC-26-151}

\maketitle


\section{Introduction}
\label{sec:intro}

Most of the gravitational-wave (GW) mergers reported until now have been detected by matched-filtering searches, which filter the data with banks of template waveforms~\cite{LIGOScientific:2016dsl, LIGOScientific:2018mvr, LIGOScientific:2020ibl, LIGOScientific:2021usb, KAGRA:2021vkt, LIGOScientific:2025slb, LIGOScientific:2024elc, LIGOScientific:2021tfm, Messick:2016aqy, Sachdev:2019vvd, Usman:2015kfa, Nitz:2017svb, Andres:2021vew, Venumadhav:2019tad, Venumadhav:2019lyq, Olsen:2022pin, Mehta:2023zlk, Wadekar:2023gea, Wadekar:2024zdq, Nitz:2018imz, Nitz:2020oeq, Nitz:2021uxj, Nitz:2021zwj, Chia:2023tle, Kumar:2024bfe}. As the number of detections grows, so does the chance of observing a gravitationally lensed signal. A fraction of the detected mergers is expected to be strongly lensed by intervening galaxies and clusters~\cite{Wang:1996as, Ng:2017yiu, Li:2018prc, Oguri:2018muv, Smith:2017mqu, Robertson:2020mfh, Smith:2022vbp, Wierda:2021upe, Xu:2021bfn, Shan:2026beq}, and signals can also pass close to stars, compact objects or intermediate-mass black holes, which act as microlenses~\cite{Lai:2018rto, Christian:2018vsi, Jung:2017flg, Diego:2019lcd, Meena:2019ate, Cheung:2020okf, Mishra:2021xzz, Shan:2022xfx, Meena:2022unp, Shan:2023ngi, Yeung:2021chy, Mishra:2023ddt, Zumalacarregui:2026uqs}. A detection of lensed GWs would enable, e.g., sub-arcsecond localization of the source, tests of the propagation speed of GWs, cosmological measurements, and constraints on compact dark matter~\cite{Hannuksela:2020xor, Fan:2016swi, Collett:2016dey, Baker:2016reh, Liao:2017ioi, Jana:2022shb, Ezquiaga:2020dao, Goyal:2020bkm, Basak:2021ten, Urrutia:2021qak, Tambalo:2022wlm, GilChoi:2023ahp, Zumalacarregui:2024ocb}; see Refs.~\cite{Grespan:2023cpa, Smith:2025axx} for reviews. No lensed GW signal has been confidently identified so far~\cite{Hannuksela:2019kle, Liu:2020par, Dai:2020tpj, LIGOScientific:2021izm, Janquart:2023mvf, LIGOScientific:2023bwz, LIGOScientific:2025cwb, Chakraborty:2025maj}, although the recent event GW231123~\cite{LIGOScientific:2025rsn} has been discussed as a microlensing candidate~\cite{LIGOScientific:2025cwb, Goyal:2025eqo, Chan:2025kyu, Shan:2025dcd, Chakraborty:2025pxt, Cheung:2026pky}.

When the gravitational wavelength is short compared with the Schwarzschild
radius of the lens, lensing is geometrical: the lens produces a discrete set of
images, each a rescaled, delayed and phase-shifted copy of the unlensed
signal~\cite{Nakamura:1997sw, Takahashi:2003ix}.
In the opposite limit, wave-optics effects become important~\cite{Ohanian:1974ys, Deguchi:1986zz, Nakamura:1999uwi, Tambalo:2022plm, Villarrubia-Rojo:2024xcj}. The relative phase of the images is set by their Morse index~\cite{Dai:2017huk, Ezquiaga:2020gdt}, and the delay between them by the lens mass. For microlenses of $10^2$--$10^5\,\msun$, the delay is milliseconds to a second, i.e., shorter than the duration of the signal in the detector band. The images then overlap and interfere, which produces a characteristic beating pattern in the waveform~\cite{Christian:2018vsi, Cheung:2020okf}. For galaxy lenses, the delay is minutes to months, and the images arrive as separate events~\cite{Haris:2018vmn, Oguri:2018muv, Wierda:2021upe}.

Current searches filter the data only with unlensed templates. For overlapping images, this has two costs. First, an unlensed template recovers only part of the signal-to-noise ratio (SNR) of the lensed signal~\cite{Mishra:2021xzz, Mishra:2023ddt, Gholap:2026}. Second, and more importantly, the interference pattern makes the lensed signal fail the signal-consistency tests that searches use to down-rank instrumental glitches~\cite{Allen:2004gu, Babak:2012zx, Usman:2015kfa, Venumadhav:2019tad}. Microlensing is therefore currently probed only by follow-up analyses of events already found by unlensed searches~\cite{Wright:2021cbn, Basak:2021ten, LIGOScientific:2023bwz, Janquart:2023mvf, Seo:2021psp, Seo:2025dto}. The alternative is to include lensed templates in the search. Ref.~\cite{Gholap:2026} (hereafter G26) recently constructed the first template bank targeted at microlensed signals, for isolated point-mass lenses. They showed that the geometric-optics approximation can efficiently build such a bank. Their bank is roughly a Cartesian product of the unlensed bank and $\sim4000$ lensed templates, which reduces the mismatch of microlensed signals by up to a factor of $\sim\!10$. Our construction builds on both of these observations.

A related issue arises for separated images. Nearly all of the previous searches for lensed pairs required at least the brighter image to be significant on its own: either both images are detected independently and then compared~\cite{Haris:2018vmn, Hannuksela:2019kle, Liu:2020par, Dai:2020tpj, Lo:2021nae, Janquart:2021qov, Goyal:2021hxv, Barsode:2024zwv, Barsode:2025agk, Gholap:2025ccr}, or one detected image is used to perform a targeted search for a sub-threshold counterpart~\cite{McIsaac:2019use, Li:2019osa, Li:2023zdl, Goyal:2023lqf, Ng:2024ooy}. Pairs in which both images are individually sub-threshold have only been examined with a model-independent comparison of candidate lists~\cite{Chakraborty:2026yci}.

Including lensing in the templates adds two degrees of freedom: the time delay $t_d$ and relative amplitude $a$ of the images. Sampling these for every unlensed template increases the filtering cost by the number of lens points, while the extra trials can reduce sensitivity. A similar problem arises for higher harmonics, where we combined separately filtered harmonics and marginalized over their relative amplitudes and phases~\cite{Wadekar:2024zdq}. Here, we marginalize over the lens parameters using the existing unlensed SNR time series, without additional matched filtering.

In this paper, we show how we can efficiently recover the lensed signals from the output of the unlensed template bank. We use the complex SNR time series of the unlensed template, and construct from it the SNR time series of every two-image template (Sec.~\ref{sec:recombination}). We do not build separate lensed templates or run any additional matched filtering. We then construct a marginalized detection statistic over the lens parameters, which is $166$ times faster than filtering the data with the G26 bank. We show that:
$(i)$ the two-image family recovers more than $97\%$ of the SNR of wave-optics point-mass injections in the geometric-optics regime, where the unlensed template recovers as little as $67\%$, and it also substantially improves the recovery for other lens models (Sec.~\ref{sec:waveoptics});
$(ii)$ our templates restore the signal-consistency test for lensed signals, which increases the sensitive volume by a median factor of $2.3$ where the unlensed template fails the test most strongly (Sec.~\ref{sec:sens});
$(iii)$ the same identity enables a coherent search for pairs of separated images which are each too faint to be detected alone. Near fold caustics, this increases the sensitive volume by a factor of $1.8$--$2.6$ over a search that requires both images to be significant, and by $1.2$--$1.6$ over a targeted counterpart search (Sec.~\ref{sec:faint}).

\section{Detection statistics for the two-image family}
\label{sec:recombination}

\subsection{The two-image family}
\label{sec:family}

An isolated point-mass lens produces a minimum and a saddle image, with a relative Morse phase of $\pi/2$~\cite{Dai:2017huk, Ezquiaga:2020gdt}. We describe the pair by its delay $t_d$ and amplitude ratio $a\equiv\sqrt{|\mu_-|/\mu_+}$, where $\mu_+>0$ and $\mu_-<0$ are the signed magnifications (Appendix~\ref{app:pointmass}).

Inspired by this case, we construct the
unit-norm two-image template with its leading image at an arbitrary time $t$:
\be
h_{\Lensed,t}(\theta) = \frac{h_t + a\, i^{\,q}\, h_{t+t_d}}{\sqrt{N(\theta)}},
\label{eq:family}
\ee
with $\theta = (t_d,a)$, $h_t$ the unit-norm unlensed template with
coalescence at $t$, and
\be
N(\theta) = 1 + a^2 + 2a\,\mathrm{Re}\!\left[i^{-q}\,C(t_d)\right]
\label{eq:norm}
\ee
the squared norm of the numerator, written in terms of the template
autocorrelation
\be
C(\tau) \equiv (h_\tau\,|\,h_0) = 4\int_0^\infty
\frac{|\tilde h(f)|^2}{S_n(f)}\, e^{2\pi i f\tau}\,\dd f .
\label{eq:autocorr}
\ee
Here $(a|b)$ is the usual noise-weighted inner product.
The factor $i^{\,q}$ is the relative Morse phase: an image
of index $n$ carries $e^{-in\pi/2}$, so any ordered pair has
$q\in\{0,1,2,3\}$. For reference, the point-mass case (a minimum followed by a saddle) is $q=1$. 

We build on the ordinary complex SNR time series of the unlensed template,
$z_1(t) \equiv (h_t\,|\,d)$. From the same time series, we construct the SNR time series of the two-image template as
\be
\left|z_\Lensed(t;\theta)\right|^2 =
\frac{|z_1|^2 + a^2|z_{1s}|^2
+ 2a\,\mathrm{Re}\!\left[i^{\,q}\, z_1 z_{1s}^{*}\right]}{N(\theta)},
\label{eq:recombine}
\ee
with $z_{1s}(t) \equiv z_1(t+t_d)$. The lens enters only through $a$, $t_d$
and the integer $q$.
We therefore do not build separate lensed templates or run any additional matched filtering.
Our formalism as presented is limited to two images, but could be generalized linearly to $n$ images if needed.


\subsection{A marginalized statistic}
\label{sec:statistic}

To build a detection statistic, we first define a grid of possible lens parameters $\theta_k=(t_d,a)_k$ and their prior weights $w_k$ (normalized to $\sum_k w_k = 1$). We then construct a marginalized statistic as
\be
S_{\rm marg}(t) = \ln \sum_k w_k
\exp\!\left[\tfrac{1}{2}|z_\Lensed(t;\theta_k)|^2\right].
\label{eq:ssoft}
\ee
For reference, the ordinary unlensed statistic is $S_\UL(t) = \tfrac{1}{2}|z_\UL(t)|^2$, and its
lensed counterpart maximized over the grid is
$S_{\max}(t) = \max_\theta \tfrac{1}{2}|z_\Lensed(t;\theta)|^2$. 


We currently profile out the image pairing $q$, signal amplitude, and coalescence phase by maximizing over them, but one could in principle marginalize over them as well.

Our prior is log-uniform in the redshifted lens mass $\MLz$ over $[10^2,10^5]\,\msun$ with an area prior
$p(y)\propto y$ over $y\in[0.01,2]$, where $y$ is the source offset in Einstein-radius units (i.e., sources distributed uniformly in area behind a point-mass lens), which corresponds to
$1~\mathrm{ms}\lesssim t_d\lesssim 1$~s. We discuss longer delays, where the images are resolved in time, in Sec.~\ref{sec:faint}.
This range is not intrinsic to the construction; a search expecting a
different lens population could change it.
We calibrate each statistic on
\emph{its own} noise background, since $S_{\max}$ and $S_{\rm marg}$ search
more templates than $S_\UL$ and therefore carry a larger look-elsewhere penalty (see Sec.~\ref{sec:volume}).

\subsection{Computational cost}
\label{sec:computational_cost}


We now compare the cost of our approach with that of the G26 bank~\cite{Gholap:2026}, which contains $4032$ explicitly constructed lensed templates for each unlensed binary configuration. Using this bank requires filtering the data with each of these templates one by one. In our case, we do not perform any matched filtering beyond that of the unlensed search.
Simply evaluating the lensed
statistic in Eq.~\eqref{eq:recombine} over every sample of the data is already $\sim 10$ times faster than filtering with the lensed templates one by one. We make one important addition. Because
$|A+B|^2\le2(|A|^2+|B|^2)$, every lens point obeys
$|z_\Lensed(t;\theta)|\le K\max\{|z_1(t)|,|z_1(t+t_d)|\}$ with
$K=\max_\theta\sqrt{2(1+a^2)/N(\theta)}$, which is $K=1.51$ on our lens grid. A
crossing of a floor $\rho_{\rm fl}$ therefore requires at least one of the two
images to be individually loud, $|z_1|\ge\rho_{\rm fl}/K$, so
Eq.~\eqref{eq:recombine} need only be evaluated at those times.
One could ideally set $\rho_{\rm fl}$ to the expected SNR threshold of the search;
even the lower floor $\rho_{\rm fl}=5$ reduces the number of evaluations of
Eq.~\eqref{eq:recombine} to $1.5\%$ of the time $\times$ lens-point total. This results in an overall speed-up of
$166$ over explicit filtering with the G26 bank. We discuss the timing and evaluation cost further in Appendix~\ref{app:cost}.


\begin{figure*}[!tb]
\includegraphics[width=\textwidth]{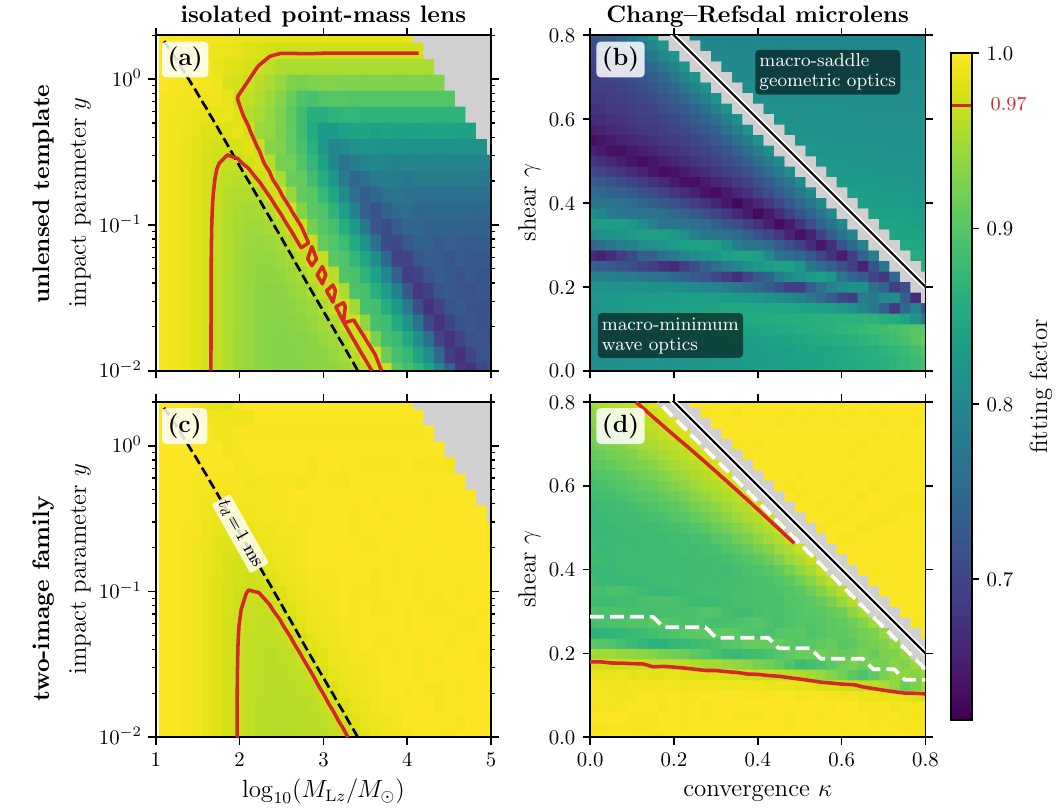}
\caption{Fraction of the SNR of lensed injections recovered by the unlensed template (top row) and by our two-image templates from Eq.~\ref{eq:family} (bottom row), for an isolated point-mass lens (left column) and for a microlens embedded in the convergence $\kappa$ and shear $\gamma$ of a macromodel (Chang--Refsdal lens, right column). In all panels, we inject exact wave-optics signals of a $M_{\rm tot}=50\,\msun$ source binary (except above the black critical line in the right panels corresponding to the macro-saddle region). See Sec.~\ref{sec:changrefsdal} for the lens parameters of the right column. The colour shows the fitting factor, where we maximize over the time delay and relative amplitude of the secondary image.
The red contour shows the $0.97$ minimal match. We fix the intrinsic binary parameters here; maximizing over them changes the results only slightly (Appendix~\ref{app:intrinsic}).
\textbf{Left column:} In the geometric-optics regime (to the right of the dashed line $t_d=1$~ms), our two-image search is optimal, while losing accuracy slightly in the wave-optics regime. 
\textbf{Right column:} Our two-image search recovers nearly perfect match in the two-image limit; the degradation occurs in the wedge region corresponding to the four-image case (see Fig.~\ref{fig:appcr}). Our two-image templates improve recovery over the unlensed template across both lens models, including where the two-image approximation degrades. 
See Sec.~\ref{sec:waveoptics} and Appendices~\ref{app:crwave} and~\ref{app:otherlenses} for details and tests on other lens models.}
\label{fig:effectualness}
\end{figure*}

\section{Signal recovery for different lenses}
\label{sec:waveoptics}

Our template family was derived for an isolated point mass in the geometric optics limit,
while astrophysical lenses are neither isolated nor axisymmetric, and wave-optics effects can be important. We test both cases by computing the fitting factor of our templates on injections from such lenses, i.e., the largest
fraction of the injection's SNR recovered by any template, maximized over the lens parameters $\theta$ of the grid in Eq.~\eqref{eq:ssoft}.\footnote{We compute the wave-optics amplification factor of the point mass analytically, and those of the other lenses with \texttt{GLoW}~\cite{Villarrubia-Rojo:2024xcj} (\url{https://github.com/glow-astro/GLoW}), which also reproduces our point-mass amplification factor.}
Appendix~\ref{app:density}
gives the lens grid that we use in this section.

\subsection{An isolated point mass, in full wave optics}
\label{sec:pointmass}

For a point-mass lens near alignment, diffraction becomes important when the phase difference associated with the image time delay satisfies \(2\pi f_{\mathrm{GW}} t_d\lesssim1\), where $f_{\mathrm{GW}}$ is the detected GW frequency and $2\pi f_{\mathrm{GW}}t_d\simeq25\,(\MLz/10^3\,\msun)(f_{\mathrm{GW}}/100~\mathrm{Hz})\,y$ for $y\lesssim1$.
Figure~\ref{fig:effectualness} shows the fitting factor for $M_{\rm tot}=50\,\msun$ (the grey cells in the left column have delays beyond our search grid); we obtain similar results for
$M_{\rm tot}=11\,\msun$ and $100\,\msun$ (Appendix~\ref{app:crwave}).
For $t_d\gtrsim1$~ms, the lens is in the geometric-optics regime. Unsurprisingly, our
two-image templates surpass the $0.97$ minimal match over this entire region, while
the ordinary unlensed search does so over only $\sim\!20\%$ of it
and falls to $0.67$ in the worst cell. Maximizing over the intrinsic binary parameters, as a real search would, changes these numbers only slightly (Appendix~\ref{app:intrinsic}).

For $t_d\lesssim1$~ms, our geometric-optics templates start to fail, but they still improve the fitting factor over the unlensed search. This is because maximizing over the delay of the two images (a linear phase term) and their relative amplitude partially absorbs the frequency dependence of the amplification factor. We discuss this in more detail in Appendix~\ref{app:crwave}.



\subsection{A microlens inside a strongly lensed image}
\label{sec:changrefsdal}

Astrophysical microlenses are typically not isolated, but are embedded in the convergence $\kappa$ and shear $\gamma$ of a galaxy or cluster.
This is the Chang--Refsdal lens model~\cite{Chang:1979zz,An:2006bq,Goyal:2025eqo} of the right
column of Fig.~\ref{fig:effectualness}. There, we vary $\kappa$ and $\gamma$ and fix the microlens mass to $\MLz=3000\,\msun$, the source position to $y=0.3$ in units of the microlens Einstein radius, and the shear position angle to $\phi_\gamma=45^\circ$. We also repeated the analysis for $\MLz=300$ and $10^4\,\msun$, and for $21$ combinations of $y\in[0.05,1.2]$ and $\phi_\gamma\in\{15^\circ,45^\circ,75^\circ\}$, and found similar results: in the macro-saddle region, the median fitting factor of our two-image templates is between $0.983$ and $0.999$ in all cases.

The critical line $\gamma=1-\kappa$ (black line in Fig.~\ref{fig:effectualness}) separates a macro-minimum from a
macro-saddle. The diagonal structure of the right column of Fig.~\ref{fig:effectualness} is due to the mass-sheet degeneracy~\cite{Schneider:1992bmb, Schneider:2013sxa}, under which the lens depends on $\kappa$ and $\gamma$ only through the reduced shear $\gamma/(1-\kappa)$, so that, e.g., the critical line is $\gamma/(1-\kappa)=1$. In the macro-saddle region, both surviving images are \emph{saddles}.
Their Morse factors are then equal, so the relative coefficient in Eq.~\eqref{eq:family} is
real and positive ($q=0$), and using the point-mass value $q=1$ degrades the fitting factor significantly. We therefore maximize over $q\in\{0,1,2,3\}$; one could also marginalize over it with an astrophysical
prior.

The fitting factor is lower in the macro-minimum half of the plane, because the injections there are wave-optical and include a four-image region
that no two-image family can represent completely. Our two-image family
still recovers substantially more of the signal there than the unlensed
template does (Appendix~\ref{app:crwave}).

\subsection{Other lens models}
\label{sec:iso}

We also test our two-image templates on two extended lens models, the singular isothermal sphere (SIS) and the cored isothermal sphere (CIS). The SIS also has a point caustic, and our two-image templates perform essentially as for the point mass, while the unlensed search falls below the $0.97$ minimal match over most of the plane (Fig.~\ref{fig:appsis}). The CIS produces three images, a central maximum in addition to the minimum and the saddle. Our two-image templates therefore recover less of the signal, but still significantly more than the unlensed search (Fig.~\ref{fig:appcored}). We discuss both cases in Appendix~\ref{app:otherlenses}.

\begin{figure*}[!tb]
\includegraphics[width=\textwidth]{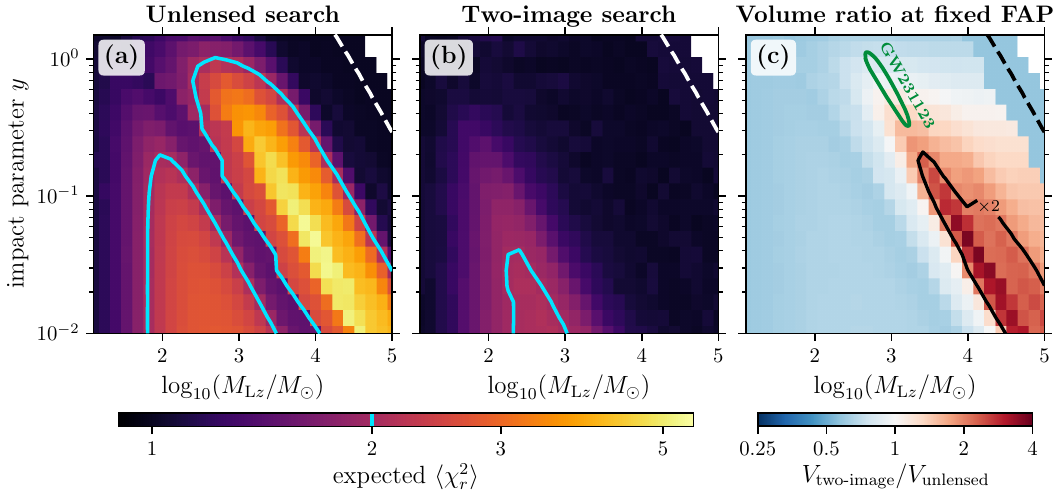}
\caption{Overlapping lensed signals can fail the standard signal-consistency test; our two-image templates restore consistency. We inject exact wave-optics point-mass signals of a $M_{\rm tot}=50\,\msun$ binary, scaled to an optimal SNR of $20$, and show the expected reduced power-$\chi^2$ statistic $\langle \chi_r^2 \rangle$ in Gaussian noise (Sec.~\ref{sec:chisq}). Both templates are maximized over the aligned-spin intrinsic parameters of the binary (Appendix~\ref{app:intrinsic}). $\langle \chi_r^2 \rangle \approx1$ for a signal that matches the template, and searches down-rank triggers with larger values.
\textbf{(a)} Using the unlensed template, the median $\langle\chi^2_r\rangle$ is $2.0$; such triggers would be strongly down-ranked in a search.
\textbf{(b)} Using the best-fit two-image template, the median falls to $1.04$.
\textbf{(c)} Ratio of the sensitive volumes at the same false-alarm probability, $10^{-3}$.
 The black contour encloses significant gains above $\times2$ (the largest being $\times3.7$). The mild decrease in other regions is because lensing barely distorts the waveform in those regions, but we pay the additional trials factor penalty for the extra lens parameters. See Sec.~\ref{sec:sens} and Appendix~\ref{app:chisq} for details.
 For reference, we also show the $90\%$ credible region of the posterior of GW231123 from Ref.~\cite{Cheung:2026pky} assuming its lensing interpretation.}
\label{fig:chisq}
\end{figure*}

\section{Signal consistency and sensitivity for overlapping lensed signals}
\label{sec:sens}

State-of-the-art GW searches do not rank their triggers by signal-to-noise ratio alone, but by an
SNR re-weighted by different types of signal-consistency tests.
For a lensed signal, this matters in a
way the recovery fractions of Sec.~\ref{sec:waveoptics} do not capture: an
overlapping-image signal is not a clean chirp, so it is down-ranked as though
it were a glitch rather than merely losing SNR to mismatch.
Ref.~\cite{Chan:2024qmb} found this effect in the GstLAL search, where the signal-consistency test lowers the detection efficiency of lensed signals with wave-optics effects from $90\%$ to below $1\%$, and argued that template banks should be extended to include lensing; here we quantify how much of this loss a two-image search recovers.

\subsection{Why a lensed signal fails the consistency test}
\label{sec:chisq}

The power that an unlensed template misses in a lensed signal is not spread evenly across the band.
The two lensed images interfere, multiplying the unlensed waveform by
$|1+a\,i^{\,q}e^{- i2\pi f t_d}|$, which alternately adds and removes power on
a frequency scale $1/t_d$.
As a realistic example, we test this with PyCBC's power $\chi^2$
\cite{Allen:2004gu}, which splits the template band into $p=16$ pieces of equal
expected contribution and compares each piece against its share, and write $\langle\chi^2_r\rangle$ for the value
expected once Gaussian noise is added (Appendix~\ref{app:chisq}).

The dashed line in the top right of each panel of Fig.~\ref{fig:chisq} marks where the delay equals the duration of the signal in the detector band, $t_d=1.16$~s for this binary. To its right, the two images no longer overlap and arrive as separate signals, which we discuss in Sec.~\ref{sec:faint}. All the delays in our searched domain ($t_d\le0.5$~s) lie to its left, so the two images overlap. Since a real search would pick the template in its bank that best fits the data, in this section we maximize both the unlensed and the two-image templates over the aligned-spin intrinsic parameters of the binary (Appendix~\ref{app:intrinsic}).
Figure~\ref{fig:chisq}(a) shows the result for the unlensed template: the
median $\langle\chi^2_r\rangle$ over the searched delay domain is $2.0$ and
exceeds $2$ over half of it, which is enough to demote a real signal below a
marginal candidate. Panel~(b) shows the same test for the best-fit two-image template.
The median $\langle\chi^2_r\rangle$ falls to $1.04$, with a significant improvement also in the wave-optics regime. We give the details of the consistency test in Appendix~\ref{app:chisq}.

\subsection{Sensitive volume analysis}
\label{sec:volume}

A naive estimate of the sensitive-volume gain is the cube of the ratio of the recovered SNRs,
$(\mathrm{FF}_\Lensed/\mathrm{FF}_\UL)^3$. 
A more accurate estimate includes the consistency test, and compares the two searches at a fixed false-alarm probability. The latter point is important because the extra lens degrees of freedom add trials.
We rank each trigger not by $\rho$ but by the
re-weighted $\hat\rho=\rho\left[\tfrac{1}{2}\left(1+(\chi^2_r)^{3}\right)\right]^{-1/6}$
for $\chi^2_r>1$, and $\hat\rho=\rho$ otherwise \cite{Usman:2015kfa}.
Since the $\chi^2$ excess of a mismatched signal grows as $\rho^2$, the demotion grows with loudness. See Appendix~\ref{app:chisq} for further details.

Figure~\ref{fig:chisq}(c) compares the sensitive volumes of the unlensed and the two-image searches at
$\mathrm{FAP}=10^{-3}$. The median volume gain across the plane is only $\times0.92$ (unweighted over the cells; Appendix~\ref{app:chisq}),
but the gain is concentrated where the lensed signal clearly fails the consistency test. For example, where $\langle\chi^2_r\rangle>4$, the median gain is $\times2.3$, and the largest
is $\times3.7$. Note that one could restrict the lens prior to the region where lensing distorts the waveform: where the volume gain is below one, the waveform is barely
distorted by lensing, but one still pays the extra trials factor penalty.


\begin{figure}[!tb]
\includegraphics[width=\columnwidth]{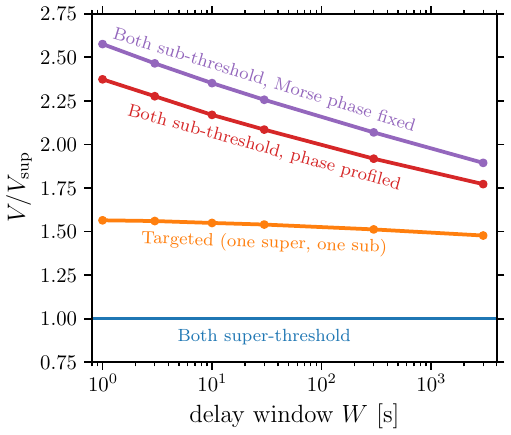}
\caption{
Our two-image method is not just useful for detecting overlapping microlensed images but also significantly increases the sensitive volume $V$ for separated strongly-lensed images.  
We compare $V$ for three strategies of Eq.~\eqref{eq:horizons} for detecting lensed pairs near a fold caustic, relative to requiring both images to be super-threshold, i.e., individually significant enough to enter a detection catalog, as a function of the window $W$ of time delays searched. All strategies are compared at the same false-alarm rate, in a Monte Carlo simulation. Our search detects pairs in which neither image is individually significant, and has the largest volume for every window, more so when the Morse phase of the two images is fixed. See Sec.~\ref{sec:faint} for details.}
\label{fig:strategies}
\end{figure}

\section{Improved search for separated lensed signals}
\label{sec:faint}

For separated images, the same construction applies with $C(t_d)\simeq0$ in Eqs.~\eqref{eq:recombine} and~\eqref{eq:norm}. Delays of minutes to months make explicit lensed-template filtering expensive, but require only combining the unlensed SNR time series at the two arrival times.

\subsection{Comparison of three strategies to detect lensed pairs}
\label{sec:strategies}

Nearly all previous searches for lensed pairs required at least the brighter image to be significant on its own, either by pairing two super-threshold detections or by a targeted search for the sub-threshold counterpart of one~\cite{McIsaac:2019use, Ng:2024ooy}. There is a complementary case in which \emph{both} images in the pair are individually sub-threshold, but the pair as a whole is significant. No existing analysis covers this case with a modeled, coherent statistic. The closest analogue in the literature is a model-independent cross-correlation of the top sub-threshold candidates from GWTC-4~\cite{Chakraborty:2026yci}, which does not exploit the two-image structure.

We compare three strategies: $(i)$ two independently significant images; $(ii)$ a targeted search for a fainter counterpart to one significant image; and $(iii)$ our pair search, which allows both images to be individually sub-threshold. Writing $\rho_\mathrm{UL}$ for the unlensed source amplitude, the images have amplitudes $\rho_\mathrm{UL}\sqrt{\mu_+}$ and $\rho_\mathrm{UL}\sqrt{|\mu_-|}$, and the detection conditions are
\begin{equation}
  \begin{split}
    &\text{both super-threshold:}\  \rho_\mathrm{UL}\sqrt{|\mu_-|} > \rho_{\mathrm{th}}\\
    &\text{one super}+\text{one sub:}\  \rho_\mathrm{UL}\sqrt{\mu_+} > \rho_{\mathrm{th}};\ \rho_\mathrm{UL}\sqrt{|\mu_-|}> \rho_\mathrm{th}^\mathrm{sub}\\
    &\text{both sub-threshold:}\  \rho_\mathrm{UL}\sqrt{\mu_++|\mu_-|} > \rho_\mathrm{th}^\mathrm{comb};\\
    &\qquad\qquad\qquad\quad\ \rho_\mathrm{UL}\sqrt{|\mu_-|}> \rho_\mathrm{sec}
  \end{split}
  \label{eq:horizons}
  \end{equation}

A search for pairs must also reject single unlensed events, which can pair up with noise to produce lensing candidates.
 We therefore also require both images to exceed a common threshold $\rho_{\rm sec}$, and we calibrate both on a background of Gaussian noise and unlensed events (Appendix~\ref{app:pairbackground}). Note that unlensed events also produce false pairs for the both-super-threshold strategy, when two unrelated detections are compatible within the window; we eschew estimating this as it requires a more detailed model of the population and thus our volume increase estimates are conservative.


 We assume sources uniform in luminosity volume, so a strategy that detects a pair down to an unlensed SNR $\rho_{\rm UL,min}$ has sensitive volume $V\propto\langle\rho_{\rm UL,min}^{-3}\rangle$, averaged over the lens population. 
The strategies recover different SNRs and incur different trials factors, so we compare them at the same expected number of false alarms, $F=0.1$ per observing run. The targeted search examines delays around catalogue detections, whereas our search examines pairs throughout the unlensed search output. We set the thresholds accordingly and choose $\rho_{\rm sec}$ and $\rho_{\rm th}^{\rm comb}$ together to maximize the volume for the target population (Appendix~\ref{app:pairbackground}). For nearly equal images, $\rho_{\rm th}^{\rm comb}\simeq\sqrt{2}\,\rho_{\rm sec}$.

\subsection{Volume gains near fold caustics}
\label{sec:faintgain}

Figure~\ref{fig:strategies} shows the sensitive volumes of the three strategies for pairs near a fold caustic, from a Monte Carlo simulation of a fold population with amplitude ratios $0.9\le a\le1$, including Gaussian noise and the antenna pattern of a single detector (Appendix~\ref{app:caustics}). Our search gains a factor of $1.8$ over requiring both images to be super-threshold for $W=50$~min and $2.4$ for $W=1$~s, which rises to $2.6$ when the Morse phase is imposed. The targeted search gains $1.5$--$1.6$, nearly independently of the window, because its brighter image must still cross $\rho_{\rm th}$, which does not depend on $W$.

These gains come from the geometry of a fold. There, two images of opposite parity ($q=1$) merge with equal magnification, and the most magnified pairs are also the most nearly equal and the closest in time, with $t_d\propto\mu^{-3}$ (Appendix~\ref{app:caustics}). For two equally bright images, the two conditions of our search in Eq.~\eqref{eq:horizons} coincide when $\rho_{\rm th}^{\rm comb}=\sqrt{2}\,\rho_{\rm sec}$, so for $W=50$~min each image needs only $\rho_{\rm sec}=6.76$ instead of $\rho_{\rm th}=8.20$, i.e., a factor of $(8.20/6.76)^3=1.78$ in volume. Since the most magnified pairs have the shortest delays, they can be searched for with a shorter window, which reduces $N_{\rm delay}$ and therefore $\rho_{\rm sec}$, so the gain grows as $W$ shrinks. Over such short delays the antenna pattern does not rotate between the two images, so their relative phase is the Morse phase $i^{\,q}$ alone, and imposing it lowers the background of the pair statistic further (Appendix~\ref{app:pairbackground}).

Note that the gain reduces for asymmetric images. For $a\lesssim0.8$, the fainter image sets the horizon of both our search and the targeted search, since $\rho_{\rm sec}\simeq\rho_{\rm th}^{\rm sub}$, and the two searches then have nearly the same volume. This is the case for an isolated point mass, whose two images have $0.38\le a\le0.84$ in the strong-lensing regime ($y\le1$): there the two searches have the same volume to within $3\%$ (Appendix~\ref{app:pointmass}). Note however that nearly equal pairs are astrophysically likely, e.g., inside the caustic of a compact lens embedded in a galaxy (Sec.~\ref{sec:changrefsdal}), where the two brightest images have a mean amplitude ratio of $0.82$--$0.94$ (Appendix~\ref{app:caustics}).

These volume ratios are robust: 
a two-detector network changes them by at most $13\%$ (Appendix~\ref{app:pairbackground}), and wider windows, as needed for delays of days to months, cost volume only logarithmically (Appendix~\ref{app:window}).

\section{Conclusions and future work}
\label{sec:conclusions}

We showed that the matched-filter output of every two-image template can be constructed from the complex SNR time series of an unlensed template, without additional matched filtering. This recovers SNR otherwise lost to lensing distortions and restores signal consistency, increasing the sensitive volume in Gaussian noise by a median factor of $2.3$ for the most strongly down-ranked signals (Figs.~\ref{fig:effectualness} and~\ref{fig:chisq}).

The same construction enables a search for separated pairs in which neither image is individually significant. Near fold caustics, it increases the sensitive volume by factors of $1.8$--$2.6$ over requiring two independent detections and $1.2$--$1.6$ over a targeted counterpart search (Fig.~\ref{fig:strategies}). The advantage over a targeted search decreases for unequal images, as for an isolated point-mass lens. This extends existing searches for resolved, repeated images~\cite{McIsaac:2019use, Li:2019osa, Li:2023zdl, Goyal:2023lqf, Ng:2024ooy, Chakraborty:2026yci, Haris:2018vmn, Hannuksela:2019kle, Dai:2020tpj, Lo:2021nae, Janquart:2021qov, Barsode:2024zwv}.

Our analysis has several simplifications which we plan to address in future work. We used a single detector with stationary Gaussian noise, and a single injected binary, for which we maximized continuously over the intrinsic parameters instead of using a full template bank. Our two-image family also degrades for $t_d\lesssim1$~ms and wherever the lens produces more than two images (Appendices~\ref{app:crwave} and~\ref{app:otherlenses}), where one could generalize Eq.~\eqref{eq:recombine} to $n$ images. Further caveats of the separated-pair search are discussed in Appendix~\ref{app:window}.

Our method only requires the complex SNR time series of each unlensed template and its autocorrelation, so it should be straightforwardly applicable to other matched-filtering pipelines, e.g.,~\cite{Messick:2016aqy, Usman:2015kfa, Andres:2021vew, Venumadhav:2019tad, Wadekar:2024zdq}. This will become more important as upgraded and next-generation detectors find many more lensed events~\cite{Ng:2017yiu, Xu:2021bfn, ET:2019dnz, Evans:2023euw}. We leave extending our statistic to multiple detectors, beyond the estimate of the trials factors in Appendix~\ref{app:pairbackground}, and running it on real LIGO--Virgo--KAGRA data~\cite{LIGOScientific:2014pky, VIRGO:2014yos, KAGRA:2020tym} to future work.

The codes associated with this paper are publicly available, together with the data and scripts needed to reproduce all of its figures.\footnote{\url{https://github.com/JayWadekar/lensed-gw-unlensed-templates}}

\emph{Note added:} While this paper was in preparation, a preprint
\cite{Chan:2026asym} appeared which uses a two-image interference model
to show that overlapping lensed signals fail standard signal-consistency
tests, in qualitative agreement with Sec.~\ref{sec:sens}, and
proposes a time-domain asymmetry statistic to recover such signals
without lensed templates. Here we instead restore the consistency test
with the two-image template selected by our statistic, and quantify the
gain in sensitive volume at fixed false-alarm probability. Our other
results, i.e., the marginalized two-image statistic, the SNR recovery
for different lens configurations, and the separated-image search, do
not overlap with Ref.~\cite{Chan:2026asym}.

\begin{acknowledgments}
We thank Sudhir Gholap, Sanjeev Dhurandhar, Shasvath Kapadia and Aaron Zimmerman for useful discussions.
We gratefully acknowledge the Texas Advanced Computing Center (TACC) at The
University of Texas at Austin for providing computational resources under
allocations PHY26026 and PHY26029, and the computing staff of the Institute for
Advanced Study for their support.
This work used large language models (LLMs), including Claude
Opus~5.5 and~5, Claude Fable~5.1 and ChatGPT Astra, to assist with literature review, coding,
running scripts, documentation, and refining parts of this text; all such use
was closely supervised by D.W. The authors take full responsibility for the scientific content, accuracy, and integrity of this paper.
M. Zaldarriaga acknowledges support
from the Nelson Center for Collaborative Research at the
Institute for Advanced Study and the Simons Foundation
through Award No. SFI-MPS-BH-00012593-10, and from the National Science Foundation through NSF-BSF 2207583 and NSF 2209991.
J.R. acknowledges support from the Simons Collaboration on Black Holes and Strong Gravity through grant SFI-MPS-BH-00012593-07, from the Kavli Institute for Cosmological Physics through an endowment from the Kavli Foundation and its founder Fred Kavli, and from the Jonathan M.\ Nelson Center for Collaborative Research.
M. Zumalacarregui is supported by grant RYC2024-049805-I, funded by MICIU/AEI/10.13039/501100011033 and by ESF+, by the CSIC Ayudas de Excelencia RYC-MaX 2024 programme, by the Severo Ochoa grant CEX2025-001574-S, funded by MICIU/AEI/10.13039/501100011033, and by the European Research Council through the ERC-CoG grant GLOW (101230608), funded by the European Union.

This research has made use of data, software and/or web tools obtained from the Gravitational Wave Open Science Center (\url{https://www.gw-openscience.org/}), a service of LIGO Laboratory, the LIGO Scientific Collaboration and the Virgo Collaboration. LIGO Laboratory and Advanced LIGO are funded by the United States National Science Foundation (NSF) as well as the Science and Technology Facilities Council (STFC) of the United Kingdom, the Max-Planck-Society (MPS), and the State of Niedersachsen/Germany for support of the construction of Advanced LIGO and construction and operation of the GEO600 detector. Additional support for Advanced LIGO was provided by the Australian Research Council. Virgo is funded, through the European Gravitational Observatory (EGO), by the French Centre National de Recherche Scientifique (CNRS), the Italian Istituto Nazionale di Fisica Nucleare (INFN) and the Dutch Nikhef, with contributions by institutions from Belgium, Germany, Greece, Hungary, Ireland, Japan, Monaco, Poland, Portugal, Spain.
\end{acknowledgments}

\appendix

\section{Lens grid, comparison with G26, and computational cost}
\label{app:cost}
\label{app:density}
\label{app:g26}
\label{app:conventions}

We maximize the fitting factors of Fig.~\ref{fig:effectualness}, the
sensitivity comparison of Sec.~\ref{sec:volume} and every appendix figure over
the same lens grid: $1000$ delays uniformly spaced in
$1~\mathrm{ms}\le t_d\le500$~ms by $8$ impact parameters in
$y\in[0.01,2]$, of which $5339$ points fall inside the prior's mass range
$\MLz\in[10^2,10^5]\,\msun$. Against geometric-optics two-image injections,
for which discretization is the only source of loss, this grid has a minimal
match of $0.992$. We space the delays linearly because the delay metric is
nearly flat beyond a few tens of milliseconds~\citep{Gholap:2026}, so a grid
uniform in $\log t_d$ puts its points where they are least needed. We use eight amplitude rows rather than four because they
are also the quadrature nodes of the sum in Eq.~\eqref{eq:ssoft}.

The same test reproduces the bank design of G26, against which we compare
the cost in Sec.~\ref{sec:computational_cost}: their $1008\times4$ bank
measures a minimal match of $0.975$ here, against their quoted $0.97$. We also
reproduce their headline loss, a minimum match of $0.670$ for the unlensed
template against wave-optics point-mass injections at
$M_{\rm tot}=50\,\msun$.

One convention needs care when moving Eqs.~\eqref{eq:family}
and~\eqref{eq:recombine} between papers. The
microlensing literature, G26 included, attaches $e^{+2\pi i f t_d}$ to the
trailing image, which is a delay only in the opposite Fourier convention from
the one standard in GW data analysis,
$\tilde h(f)=\int h(t)\,e^{-2\pi i f t}\,\dd t$. Used verbatim in the latter,
it places the second image \emph{before} the first. Every expression in this
paper uses the GW data-analysis convention, and we verified the signs on
zero-noise injections.


The timings behind the speed-up quoted in Sec.~\ref{sec:computational_cost}
use this G26 bank ($1008$ delays by four magnification rows) in a $16$~s segment at $4096$~Hz, on one core.
Filtering with the explicit templates takes $18.8$~s
and $4032$ inverse FFTs. Evaluating Eq.~\eqref{eq:recombine} at every sample
takes $1.90$~s and one FFT, and restricting it to the candidate samples of
Sec.~\ref{sec:computational_cost} takes $0.113$~s. The candidate restriction
is exact for $S_\UL$ and $S_{\max}$, and reproduces both the value and the
location of the maximum of the full scan.

Both approaches scale linearly with the number of lens points, so the
candidate restriction gives a constant factor, set by the fraction of samples
above the floor $\rho_{\rm fl}$ ($1.5\%$ for $\rho_{\rm fl}=5$). The remaining
fixed cost (one FFT for $z_1$, one for the autocorrelation, and selecting the
candidates) limits the speed-up on this bank to about $1400$.

\section{Calibration of the detection statistics, and the population-averaged volume}
\label{app:statistics}
\label{app:null}

Section~\ref{sec:statistic} calibrates each statistic on its own noise
background. Here we describe that calibration, and the sensitive volume it
gives without the signal-consistency test of Sec.~\ref{sec:sens}. We
calibrate $S_\UL$, $S_{\max}$ and $S_{\rm marg}$ on the same $1.2\times10^{6}$ Gaussian-noise segments of $4$~s each. The CBC template is an equal-mass $50\,\msun$ binary, and
the lens grid is the $5339$-point grid of Appendix~\ref{app:density} with the
prior of Sec.~\ref{sec:statistic}. At a false-alarm probability of $10^{-3}$
per segment, this leaves $1200$ background exceedances per statistic, so no
extrapolation is needed. We evaluate the lens grid only at the candidate
samples of Sec.~\ref{sec:computational_cost}, which is exact for $S_\UL$ and
$S_{\max}$ and conservative for $S_{\rm marg}$.

Relative to $S_\UL$, maximizing over the lens grid raises the threshold of
$S_{\max}$ by $6.35$ at this false-alarm probability, while marginalizing
raises that of $S_{\rm marg}$ by only $0.04$. Without the signal-consistency
test of Sec.~\ref{sec:sens}, and averaged over the lens prior, the sensitive
volumes relative to the unlensed search are
\begin{align}
\frac{V_{\max}}{V_\UL} &= 0.888\,[0.882,\,0.895], \nonumber\\
\frac{V_{\rm marg}}{V_\UL} &= 1.001\,[0.999,\,1.004],
\label{eq:nullvolume}
\end{align}
with bootstrap $1\sigma$ intervals over the noise realizations. Maximizing
over the grid loses $11\%$ of the volume, and marginalizing neither
gains nor loses. This null reflects the lens prior: the unlensed search loses SNR mainly at large $\MLz$ and
small $y$ (Fig.~\ref{fig:effectualness}), while the area prior
$p(y)\propto y$ puts most of its weight at large $y$, where $a$ is small and
the lensed template is nearly the unlensed one. The gain of Sec.~\ref{sec:volume}
and Fig.~\ref{fig:chisq}(c) comes from the consistency test, which down-ranks lensed signals in exactly
that corner.

The null does not depend on these choices: for other priors, a wider lens
domain, and lens grids of $239$ to $7373$ points, $V_{\rm marg}/V_\UL$ stays
within $0.003$ of unity, while $V_{\max}/V_\UL$ decreases as the grid grows,
since more grid points add trials to a maximized statistic but not to a
marginalized one. 

\section{Other CBC masses and lens models}
\label{app:lenses}

\subsection{Point mass and Chang--Refsdal lens in wave optics}
\label{app:crwave}
\label{app:morse}

\begin{figure*}[!tb]
\includegraphics[width=\textwidth]{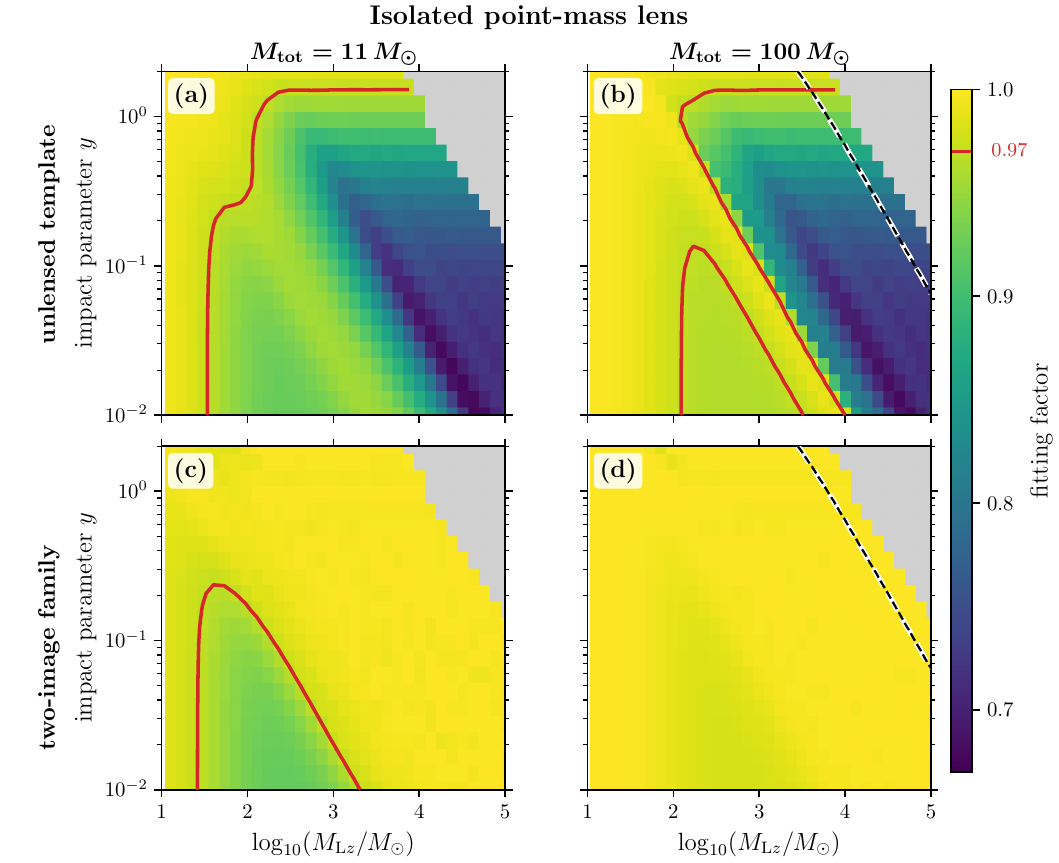}
\caption{Same as the left column of Fig.~\ref{fig:effectualness}, but for binaries with $M_{\rm tot}=11\,\msun$ (left) and $100\,\msun$ (right), and with the search grid of Appendix~\ref{app:density} (grey cells have $t_d>500$~ms). Over the searched delays $1~\mathrm{ms}\le t_d\le500$~ms, the unlensed template exceeds the minimal match in only $15\%$ and $31\%$ of the cells, while our two-image templates exceed it in every cell. The lower fitting factors at small $\MLz$ in the bottom row are at $t_d<1$~ms, outside the searched delays, where geometric optics fails. The dashed line in the right column is the chirp duration ($0.257$~s at $100\,\msun$; it is off scale at $11\,\msun$). Beyond it, the two images no longer overlap and the separated-image search of Sec.~\ref{sec:faint} applies.}
\label{fig:apppm}
\end{figure*}

This appendix expands on Secs.~\ref{sec:pointmass}
and~\ref{sec:changrefsdal}. Figure~\ref{fig:apppm} repeats the point-mass
panels of Fig.~\ref{fig:effectualness}(a, c) at $M_{\rm tot}=11$ and $100\,\msun$, on the same
grid. The two-image family clears the minimal match in every searched cell at
both masses, and the unlensed template loses about as much, and in the same
place, as at $50\,\msun$. Maximizing over $q$ changes nothing measurable, as
expected for a minimum--saddle pair. The CBC mass changes only the width of the
region where geometric optics fails, not its position.

As noted in Sec.~\ref{sec:pointmass}, our two-image templates still improve
on the unlensed search below the $t_d=1$~ms contour of
Fig.~\ref{fig:effectualness}(c), where the injections are wave-optical. The exact amplification factor approaches its
geometric-optics limit quickly with the dimensionless frequency
$w=8\pi G\MLz f/c^3$: at $y=0.3$ the fractional difference is $0.24$ at $w=1$
and $0.04$ at $w=3$. For $M_{\rm tot}=50\,\msun$ only $3.7\%$ of $\rho^2$ lies
below $w=3$ at $\MLz=10^3\,\msun$, while at $\MLz=10^2\,\msun$ most of it lies
below $w=1$, where the family fails. In between, the free $(t_d,a)$ and the
coalescence time and phase absorb the slowly varying part of the residual.

\begin{figure*}[!tb]
\includegraphics[width=\textwidth]{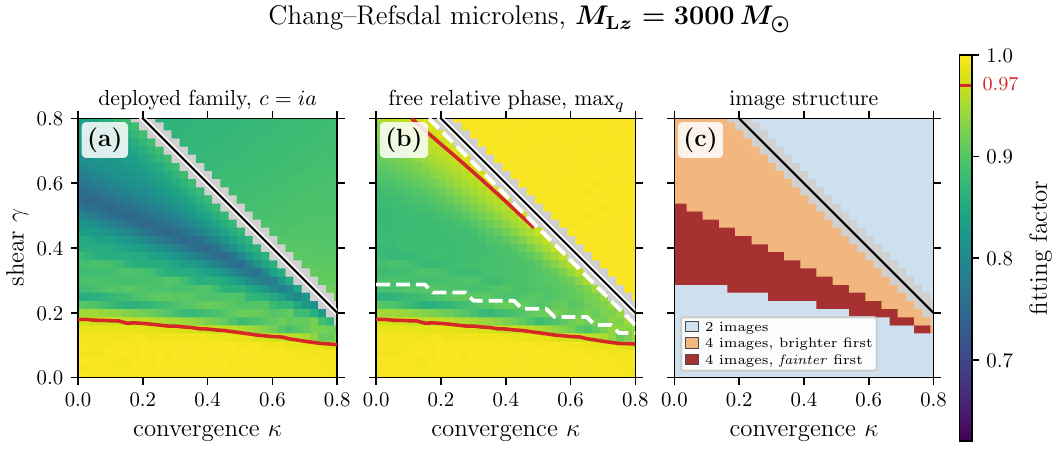}
\caption{The Chang--Refsdal lens of Fig.~\ref{fig:effectualness}(b, d) in more detail ($\MLz=3000\,\msun$, $y=0.3$, $\phi_\gamma=45^\circ$, $M_{\rm tot}=50\,\msun$), with the same colour scale and red $0.97$ contour. The solid black line is the critical line $\gamma=1-\kappa$, and we exclude the grey band around it, where the macro-magnification diverges. We use exact wave-optics injections in the macro-minimum region and exact geometric-optics injections in the macro-saddle region.
\textbf{(a)} Two-image templates with the point-mass Morse phase ($q=1$): the median fitting factor is $0.884$, and $21\%$ of the cells exceed the minimal match, none of them in the macro-saddle region.
\textbf{(b)} Maximizing over $q$: the median rises to $0.970$, and $50\%$ of the cells exceed the minimal match, including all of the macro-saddle cells.
\textbf{(c)} Number of images and their arrival order. The templates of panel (b) exceed the minimal match in $83\%$ of the $590$ two-image cells, but in only $7\%$ of the $268$ four-image cells where the brighter image arrives first, and in none of the four-image cells where the fainter image arrives first. For comparison, the unlensed template does not exceed the minimal match in any of the cells (Fig.~\ref{fig:effectualness}b).}
\label{fig:appcr}
\end{figure*}

Figure~\ref{fig:appcr} shows where our family works for the Chang--Refsdal
lens of Fig.~\ref{fig:effectualness}(b, d). In the macro-saddle region both
images are saddles and the relative Morse phase is $q=0$
(Sec.~\ref{sec:changrefsdal}). A family fixed at $q=1$ then has a floor: for
well-separated images ($C(t_d)\to0$) against a $q=0$ pair of amplitude ratio
$a_\star$, the best overlap it can reach is
\be
\max_a \frac{\sqrt{1+a^2a_\star^2}}{\sqrt{(1+a_\star^2)(1+a^2)}}
 = \frac{1}{\sqrt{1+a_\star^2}} ,
\label{eq:morsefloor}
\ee
attained at $a=0$, i.e., the best the family can do is drop the second image
and become the unlensed template. For the median $a_\star=0.754$ of this
region the floor is $0.80$, and indeed no macro-saddle cell clears the minimal
match at $q=1$. Maximizing over $q$ selects $q=0$ in every cell and reaches a
median of $0.998$. Since $q$ enters Eq.~\eqref{eq:recombine} only through the
coefficient $i^{\,q}$, this costs no extra filtering, although it adds trials
to the search, which we have not quantified.

Inside the caustic the lens makes four images, which no two-image template can
represent at any $a$ or $q$ (Fig.~\ref{fig:appcr}c). In some of these cells the
fainter image arrives first, so that $a>1$, outside our grid. Extending the
grid to $a>1$ raises their median fitting factor only from $0.887$ to $0.907$,
so the loss is due to the extra images rather than to the amplitude range of
the grid.

\subsection{Isothermal spheres and a third image}
\label{app:otherlenses}

For the lenses of Sec.~\ref{sec:iso}, we keep the
point-mass search grid and change only the injections, which are exact wave
optics computed with \texttt{GLoW}~\cite{Villarrubia-Rojo:2024xcj}, and count
only cells inside the searched domain $1~\mathrm{ms}\le t_d\le500$~ms.
For the SIS, our templates exceed the $0.97$ minimal match in at least $99.7\%$ of the searched cells at all three binary masses, against $11$--$26\%$ for the unlensed template. For the cored sphere, maximized over $q$, they do so in $99.6$--$100\%$ of the cells for $r_c\le0.15$ and in $78\%$ for $r_c=0.30$, against at most $7\%$ for the unlensed template.

\begin{figure*}[!tb]
\includegraphics[width=\textwidth]{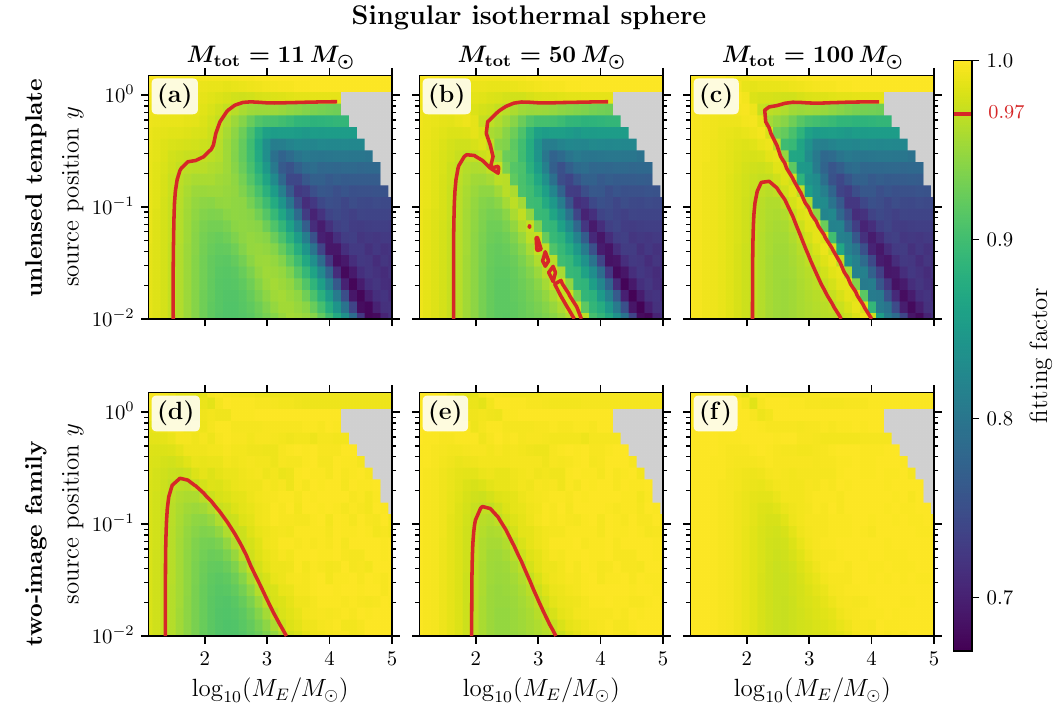}
\caption{Same as Fig.~\ref{fig:effectualness}(a, c), but for a singular isothermal sphere lens and binaries with $M_{\rm tot}=11$, $50$ and $100\,\msun$. The horizontal axis is the lens mass within the Einstein radius. We compute the exact wave-optics injections with \texttt{GLoW}~\cite{Villarrubia-Rojo:2024xcj} and search them with the same point-mass grid (grey cells have $t_d>500$~ms). Above $y\simeq1$ the lens produces a single image, so both templates recover nearly all of the signal there.}
\label{fig:appsis}
\end{figure*}

A singular isothermal sphere again produces a minimum and a saddle, so $q=1$
is already the right Morse phase, and the recovery is essentially the same as
for the point mass (Fig.~\ref{fig:appsis}).

\begin{figure*}[!tb]
\includegraphics[width=\textwidth]{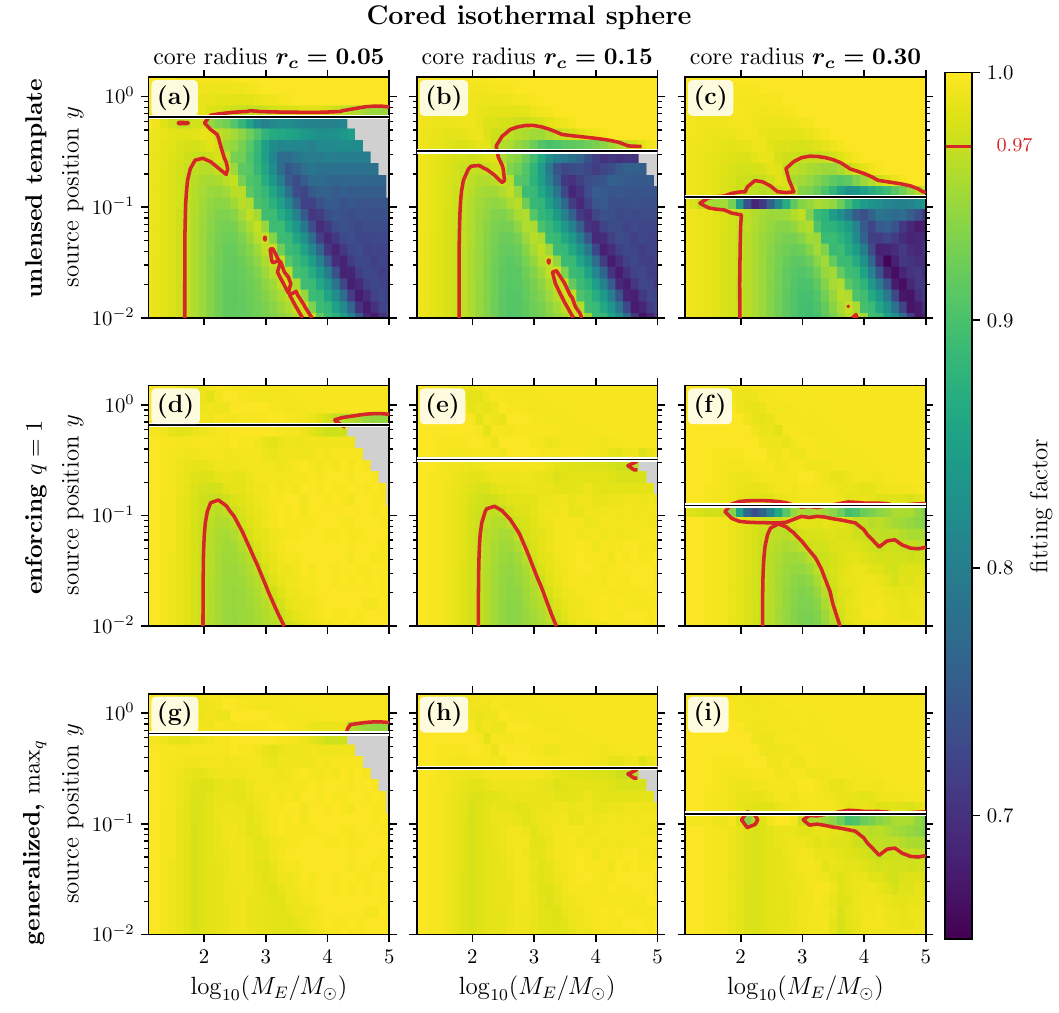}
\caption{Same as Fig.~\ref{fig:appsis}, but for a cored isothermal sphere with core radii $r_c=0.05$, $0.15$ and $0.30$, and a $M_{\rm tot}=50\,\msun$ binary. We show our two-image templates both with the point-mass Morse phase ($q=1$, middle row) and maximized over $q$ (bottom row). Below the solid black line the lens produces three images, which a two-image template cannot fully represent. A larger core pushes more of the plane to a single image or to delays outside our grid, so the number of searched cells decreases. See Appendix~\ref{app:otherlenses} for the origin of the remaining loss at $r_c=0.30$.}
\label{fig:appcored}
\end{figure*}

A cored isothermal sphere produces a third image over part of the plane, a
central maximum whose geometric-optics coefficient is real and negative, which
Eq.~\eqref{eq:family} cannot represent (Fig.~\ref{fig:appcored}). For small
cores the third image is faint, and since its amplitude ratio enters the
template normalization quadratically, the loss is small. At $r_c=0.30$ every
searched cell has three images and the brightest image arrives second, so the
family fails even when maximized over $q$. As for the Chang--Refsdal lens, the remaining loss is the extra image itself.
Recovering it would need the three-image generalization of
Eq.~\eqref{eq:recombine}, which adds cross terms but still no extra filtering.

\section{Details of the signal-consistency test and its effect on the sensitive volume}
\label{app:chisq}

This appendix gives the details of the consistency test of
Sec.~\ref{sec:chisq} and of the volume comparison of Sec.~\ref{sec:volume},
both shown in Fig.~\ref{fig:chisq}. The power $\chi^2$ is $\chi^2 = p\sum_{i=1}^{p}\left|z^{(i)}-z/p\right|^2$, where $z^{(i)}$ is the
complex SNR accumulated in the $i$th piece and
$z=\sum_i z^{(i)}$ the total, and we quote the reduced
$\chi^2_r=\chi^2/(2p-2)$. We use $p=16$ bins, so a signal that matches the template leaves $\chi^2$ at
its $2p-2=30$ degrees of freedom and $\chi^2_r=1$. The injections of
Fig.~\ref{fig:chisq} are exact wave-optics point masses, noise-free and scaled
to $\rho=20$, so the $\chi^2$ excess they produce is purely the mismatch, and
$\langle\chi^2_r\rangle=1+\text{excess}$ is the value expected in Gaussian
noise, which we verified on noise realizations.

For the two-image search we evaluate the test the way a pipeline would, for
the template that produced the trigger: we build the best-fit two-image
template on the grid (with the intrinsic parameters maximized as in Appendix~\ref{app:intrinsic}) and run the same $\chi^2$ against it on the same data.
The bins then move as well, since the equal-power bins follow $|h_\Lensed|^2$,
which carries the interference pattern. Building this one template is a
follow-up of a single trigger and does not reintroduce the per-lens-point
filtering.

The excess grows as $\rho^2$, so a single reference map extrapolates to any
distance in closed form. The background, however, has to be simulated. We rank $3\times10^{5}$
Gaussian-noise segments by the re-weighted $\hat\rho$ of Sec.~\ref{sec:volume} for each search, with the two-image
search using the $\chi^2$ of the template its own maximization selected, as on
the signal side. The thresholds are $5.47$ for the
unlensed and $6.50$ for the two-image search at $\mathrm{FAP}=10^{-3}$. That
difference is the trials factor of the lens grid, and on its own it costs a
factor $(5.47/6.50)^3=0.60$ in volume.

The medians quoted in Sec.~\ref{sec:volume} are unweighted over the cells of Fig.~\ref{fig:chisq}, which are spaced uniformly in $\log\MLz$ and $\log y$, rather than weighted by a lens population; Appendix~\ref{app:null} gives a prior-weighted comparison without the consistency test. The naive estimate $(\mathrm{FF}_\Lensed/\mathrm{FF}_\UL)^3$
has a median of $1.40$ and cannot exceed $2.9$. Including the thresholds but not the consistency test, the median
volume ratio is $0.84$. It is the
consistency test that lifts the median to $0.92$ and the loudest cells to
$\times3.7$. This is a single-detector comparison in Gaussian noise, at one PSD
and one choice of bins; a real pipeline's ranking statistic includes more than
$\hat\rho$.

\section{Maximization over the intrinsic binary parameters}
\label{app:intrinsic}

The injections of Figs.~\ref{fig:effectualness} and~\ref{fig:chisq} are a single binary, IMRPhenomD with $m_1=m_2=25\,\msun$ and no spins. A real search would filter the data with a bank of intrinsic templates, and an unlensed template with biased masses or spins could absorb part of the lensing distortion. To test this, we maximize the fitting factor over the aligned-spin, quadrupole-only parameters $(\mathcal{M},\eta,\chi_{1z},\chi_{2z})$ of IMRPhenomD for every cell, with a coarse scan followed by a Nelder--Mead optimization. For the two-image template, we keep $(t_d,a)$ at the best-fit point of the lens grid and maximize only over the intrinsic parameters. For the consistency test, the unlensed search uses whichever of two templates gives it the larger sensitive volume: the one that maximizes its SNR, and the one that maximizes its re-weighted $\hat\rho$ at $\rho=20$.

A continuous maximization is more favorable to the unlensed search than a discrete bank, while keeping $(t_d,a)$ fixed is less favorable to the two-image search, so the comparison is conservative.

For Fig.~\ref{fig:effectualness}, the effect is small. In the geometric-optics region, the median fitting factor of the unlensed template rises from $0.87$ to $0.90$, its worst cell from $0.67$ to $0.70$, and the fraction of cells above the $0.97$ minimal match from $20\%$ to $25\%$. The two-image recovery is unchanged (median $0.997$, worst cell $0.977$). We therefore show Fig.~\ref{fig:effectualness} at fixed intrinsic parameters.

For Fig.~\ref{fig:chisq}, the effect is larger, because a biased unlensed template can partially absorb the interference pattern and so pass the consistency test more easily. Maximizing over the intrinsic parameters lowers the median $\langle\chi^2_r\rangle$ of the unlensed search from $2.4$ to $2.0$ (and that of the two-image search from $1.05$ to $1.04$), and the largest volume gain from $\times7.5$ to $\times3.7$. Where $\langle\chi^2_r\rangle>4$, the median gain is $\times2.3$ with the intrinsic parameters maximized, and $\times2.6$ with them fixed. Figure~\ref{fig:chisq} and Sec.~\ref{sec:sens} quote the maximized values.

\section{Separated images behind an isolated point mass, and the delay window}
\label{app:pointmass}

An isolated point mass of redshifted mass $\MLz$, with the source at dimensionless impact parameter $y$
in units of the Einstein radius, produces two images (a minimum and a
saddle of the arrival-time surface) with signed magnifications and relative
delay
\begin{align}
\mu_\pm(y) &= \frac{1}{2} \pm \frac{y^2+2}{2y\sqrt{y^2+4}}, \label{eq:mu}\\
t_d(\MLz,y) &= \frac{4G\MLz}{c^3}
\left[\frac{y\sqrt{y^2+4}}{2}
+ \ln\!\left(\frac{\sqrt{y^2\!+\!4}+y}{\sqrt{y^2\!+\!4}-y}\right)\right],
\label{eq:td}
\end{align}
and $\mu_-<0$. Besides the delay and the magnification, the saddle image also acquires a frequency-independent phase shift of $\pi/2$ relative to the minimum, i.e., its Morse phase~\cite{Dai:2017huk, Ezquiaga:2020gdt}. We use the amplitude ratio
$a \equiv \sqrt{|\mu_-|/\mu_+}\in(0,1)$.

The fold pairs of Sec.~\ref{sec:faint} arise, e.g., inside the caustic of a compact lens embedded in a galaxy, i.e., the Chang--Refsdal lens (Appendix~\ref{app:caustics}), which is the more astrophysically motivated case. For completeness, we repeat here the comparison for an isolated point-mass lens, restricted to the strong-lensing regime, $y\le1$. We search delays of $10$--$60$~minutes, which fixes $\MLz=8.8\times10^{7}\,\msun$ and $y\in[0.17,1]$ (see below), so that the amplitude ratio runs from $a=0.84$ to $0.38$. We average over the cross-section prior $p(y)\propto y$, with the Monte Carlo simulation of Sec.~\ref{sec:faintgain}.

\begin{figure}[!tb]
\includegraphics[width=\columnwidth]{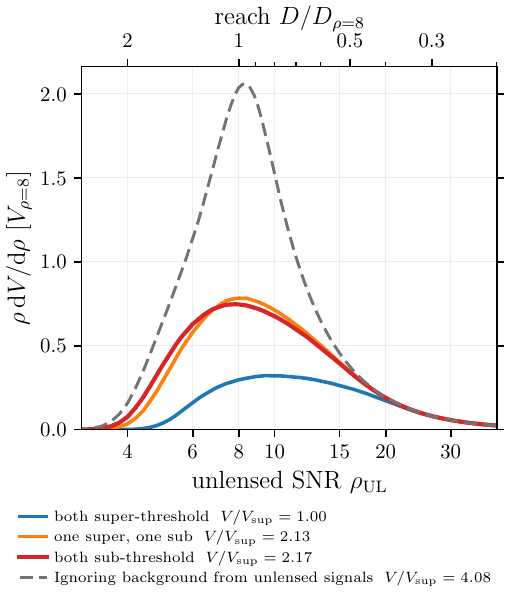}
\caption{Sensitive volume of the strategies of Sec.~\ref{sec:strategies} for an isolated point-mass lens (Appendix~\ref{app:pointmass}), as a function of the unlensed SNR $\rho_\mathrm{UL}$. We show $\rho\,\dd V/\dd\rho$, normalized to the volume within the $\rho=8$ horizon, so that the area under each curve on the logarithmic horizontal axis is proportional to the sensitive volume; the top axis shows the corresponding reach. All strategies are at $0.1$ expected false alarms per observing run, on a background of Gaussian noise and unlensed events, and the simulation includes noise and the antenna pattern. Our conservative search (red) and the targeted search (orange) have nearly the same volume, $2.2$ and $2.1$ times that of requiring both images to be super-threshold (blue). With a two-detector network, our search has $1.10$ times the volume of the targeted search (Appendix~\ref{app:pairbackground}), and it needs no catalogue detection. The comparison is conservative: we charge our search for unlensed events paired with noise, but not the both-super-threshold strategy for false pairs of unrelated detections. The dashed curve is our search calibrated on Gaussian noise alone, which does not reject unlensed events (last row of Table~\ref{tab:envelope}).}
\label{fig:apppairs}
\end{figure}

Table~\ref{tab:envelope} gives the thresholds and the volumes. The thresholds $\rho_{\rm sec}=6.31$ and $\rho_{\rm th}^{\rm comb}=10.18$ maximize the volume of our search for this population at the false-alarm budget (Appendix~\ref{app:pairbackground}). Pairs with $a<\rho_{\rm sec}/[(\rho_{\rm th}^{\rm comb})^2-\rho_{\rm sec}^2]^{1/2}=0.79$, which carry all but $3\%$ of the prior weight, are then limited by the fainter image in Eq.~\eqref{eq:horizons}, as in the targeted search with $\rho_{\rm th}^{\rm sub}=6.17$. The two searches therefore have nearly the same volume, and they also find pairs at nearly the same unlensed SNRs (Fig.~\ref{fig:apppairs}).

Without the secondary threshold, i.e., with the pair statistic calibrated on Gaussian noise alone, our search would have $4.1$ times the volume of the both-super-threshold strategy (last row of Table~\ref{tab:envelope}). However, $42\%$ of this volume comes from pairs whose fainter image cannot be distinguished from the loudest noise sample in the window, which is why we need the secondary threshold of Sec.~\ref{sec:strategies}.

\begin{table}[!tb]
\caption{Volume gains for an isolated point-mass lens (Appendix~\ref{app:pointmass}), for the three strategies of Sec.~\ref{sec:strategies} with a $10$--$60$~minute delay window (trials counts as in Appendix~\ref{app:pairbackground}). For our sub-threshold search, the thresholds are $\rho_{\rm sec}$ on each image and $\rho_{\rm th}^{\rm comb}$ on the pair, and the background also contains unlensed events. The analytic column uses the sharp horizons of Eq.~\eqref{eq:horizons}, and the simulation includes noise and the antenna pattern (Fig.~\ref{fig:apppairs}). The last row is the same search calibrated on Gaussian noise alone, without $\rho_{\rm sec}$, which does not reject unlensed events.}
\label{tab:envelope}
\begin{ruledtabular}
\begin{tabular}{lcccc}
 & & & \multicolumn{2}{c}{$V/V_{\rm sup}$} \\
strategy & $N_\mathrm{trials}$ & thresholds & analytic & sim. \\
\hline
both super-threshold      & $3.8\times10^{13}$ & $8.20$ & $1.00$ & $1.00$ \\
one super, one sub        & $1.8\times10^{7}$  & $6.17$ & $2.23$ & $2.13$ \\
both sub-threshold        & $6.7\times10^{18}$ & $6.31$, $10.18$ & $2.17$ & $2.17$ \\
\hline
No unlensed background     & $6.7\times10^{18}$ & $9.96$ & $4.04$ & $4.08$ \\
\end{tabular}
\end{ruledtabular}
\end{table}

\label{app:window}
We now describe what lens the $10$--$60$~minute window corresponds to, and
what widening it costs. Solving
$t_d(\MLz,y=1)=60$~min and $t_d(\MLz,y_{\min})=10$~min with
Eq.~\eqref{eq:td} gives $\MLz=8.8\times10^{7}\,\msun$ and $y\in[0.17,1]$. The window
therefore fixes the lens mass but keeps almost the whole range of impact
parameters, and the total magnification $\mu_++|\mu_-|$ runs from $5.8$ at
the short-delay end to $1.3$ at the long-delay end.

At $D_L\simeq1$~Gpc, this lens requires $\sim\!10^{8}\,\msun$ within its
Einstein radius of $\sim\!10^{2}$~pc, which a dark-matter halo of this mass
does not reach, but the inner regions of early-type galaxies do~\cite{Koopmans:2006iu}. The
strong-lensing optical depth is instead dominated by $\sim\!L_*$ galaxies with
$\sigma_v\sim200\,\mathrm{km\,s^{-1}}$, which produce delays of months to
years. Our window is therefore a benchmark to compare the three strategies of
Sec.~\ref{sec:strategies}, not a claim about where lensed pairs are most likely
to be found.

Widening the window changes only the trials factors of our search and of the targeted search, both of which are proportional to $N_{\rm delay}$. With Gaussian noise alone, the threshold would grow only as $\sqrt{\ln T}$, and a window of a year would cost $23\%$ of the volume. With unlensed events in the background the cost is larger, because their contribution to the false alarms also grows as $N_{\rm delay}$ and raises the secondary threshold. Table~\ref{tab:window} gives the result for the point-mass population of Appendix~\ref{app:pointmass}, re-optimizing the thresholds at each window. Covering delays up to a year costs $41\%$ of the volume, for our search and the targeted search alike. For equally bright images ($a=1$), the cost is $30\%$.

\begin{table}[!tb]
\caption{Cost of widening the delay window beyond $10$--$60$~minutes (first row), at the same expected number of false alarms ($F=0.1$ per observing run), for the point-mass population of Appendix~\ref{app:pointmass}. The trials count $N$ grows in proportion to the window, which raises the thresholds of both the targeted search ($\rho_{\rm th}^{\rm sub}$) and our search ($\rho_{\rm sec}$, re-optimized at each window), and lowers their volumes $V_{\rm one}$ and $V_{\rm sub}$, which we give relative to the first row.}
\label{tab:window}
\begin{ruledtabular}
\begin{tabular}{lccccc}
window & $N/N_{10\text{--}60}$ & $\rho_{\rm th}^{\rm sub}$ & $\rho_{\rm sec}$ & $V_{\rm one}$ & $V_{\rm sub}$ \\
\hline
$10$--$60$~min & $1$         & $6.17$ & $6.30$ & $1.00$ & $1.00$ \\
$1$ day                   & $24$             & $6.66$ & $6.75$ & $0.83$ & $0.81$ \\
$1$ week                  & $1.7\times10^{2}$ & $6.95$ & $7.05$ & $0.74$ & $0.72$ \\
$1$ month                 & $7.2\times10^{2}$ & $7.16$ & $7.20$ & $0.67$ & $0.67$ \\
$1$ year                  & $8.8\times10^{3}$ & $7.50$ & $7.55$ & $0.59$ & $0.59$ \\
\end{tabular}
\end{ruledtabular}
\end{table}

The computational cost of a wider window is negligible: our search only pairs the single-image triggers above $\rho_{\rm sec}$, about $10^{5}$ per observing run, which costs little compared with the filtering that produces them. This is an estimate from counting candidate pairs, not a timed search over an observing run.

\section{Background of the pair search}
\label{app:pairbackground}

The sensitive volume of each strategy depends on the SNR it recovers and on its number of trials, which fixes its thresholds: a significant detection requires an SNR $\simeq\sqrt{2\ln(N/F)}$ for $N$ independent trials at $F$ expected false alarms. The three strategies have different trials factors, so we compare them at the same expected number of false alarms, $F=0.1$ per observing run. An ordinary single-image search with a bank of $N_{\rm templates}$ templates examines $N_{\rm single}=T_{\rm obs}f_{\rm rms}N_{\rm templates}$ independent samples in an observing run of length $T_{\rm obs}$, where $f_{\rm rms}$ is the rate at which its filter output produces independent samples (Appendix~\ref{app:pairbackground}). Requiring both images to be super-threshold adds no further trials: each image is found by the ordinary search, and the pair is assembled afterwards from a detection list that is already significant. A search conditioned on one of the $N_{\rm catalogue}\simeq10^{2}$ super-threshold detections examines only the $N_{\rm delay}$ independent delays in the window around each, because that detection fixes the template and the sky position, where $N_{\rm delay}$ is proportional to the width $W$ of the window. Our sub-threshold search examines $N_{\rm pairs}=N_{\rm single}N_{\rm delay}$ pairs.

The thresholds of the three strategies are then approximately
\begin{equation}
\begin{split}
  \rho_{\rm th}&\simeq\sqrt{2\ln(N_{\rm single}/F)}=8.20,\\
  \rho_{\rm th}^{\rm sub}&\simeq\sqrt{2\ln(2N_{\rm catalogue}N_{\rm delay}/F)}=6.28,\\
  \rho_{\rm sec}&\simeq\sqrt{\ln(N_{\rm pairs}/F)}=6.76,\quad \rho_{\rm th}^{\rm comb}\simeq\sqrt{2}\,\rho_{\rm sec}=9.56,
\end{split}
\label{eq:thresholds}
\end{equation}
for $N_{\rm templates}=10^{4}$ and a window of $W=50$~min, for which $N_{\rm delay}=1.7\times10^{5}$ and $N_{\rm pairs}=6.7\times10^{18}$. For reference, $W=1$~s gives $\rho_{\rm th}^{\rm sub}=4.84$ and $\rho_{\rm sec}=6.14$. The factor of $2$ in $\rho_{\rm th}^{\rm sub}$ is because near a fold the order of the two images is unknown, so the window extends to both sides of the detection. The last line holds for nearly equal images, as near a fold (Sec.~\ref{sec:faintgain}), since each image of a noise pair exceeds $\rho_{\rm sec}$ with probability $e^{-\rho_{\rm sec}^2/2}$. For unequal images, $\rho_{\rm sec}$ is instead set by unlensed events with a noise partner, and is close to $\rho_{\rm th}^{\rm sub}$; in general we choose $\rho_{\rm sec}$ and $\rho_{\rm th}^{\rm comb}$ together to maximize the volume of the target population (Appendix~\ref{app:pairbackground}).

The single-image search examines $N_{\rm single}=T_{\rm obs}f_{\rm rms}N_{\rm templates}$ independent samples, where $f_{\rm rms}$ is the root-mean-square frequency of the template's SNR integrand, i.e., the rate at which its filter output produces independent samples ($f_{\rm rms}=122$~Hz for $M_{\rm tot}=50\,\msun$). For $N_{\rm templates}=10^{4}$, this gives $\rho_{\rm th}=8.20$.

The two thresholds of our pair search are the result of comparing the pair hypothesis with both the noise and the single-event hypotheses, with the amplitude and phase of each image profiled out. We set them at $F=0.1$ expected false alarms per run, on a background with two components, pairs of noise samples and unlensed events paired with a noise sample:
\be
F = N_{\rm pairs}\,P_0 + 2N_{\rm delay}\int\dd\rho\,\frac{\dd N}{\dd\rho}\,P_\rho,
\label{eq:pairbudget}
\ee
where $P_\rho\equiv P\big(\min(|z_1|,|z_2|)>\rho_{\rm sec},\ S>(\rho_{\rm th}^{\rm comb})^2\,\big|\,\rho\big)$ is the probability that a pair passes both cuts when one image is an unlensed event of SNR $\rho$ and the other is noise, and $P_0$ is the same for two noise samples. $S$ is the pair statistic, and $|z|^2$ of a noise sample follows a $\chi^2$ distribution with two degrees of freedom, while that of the event follows a non-central $\chi^2$ distribution with non-centrality $\rho^2$. The factor of $2$ is because the event can be either image of the pair. For the incoherent statistic $S=|z_1|^2+|z_2|^2$, $P_0=e^{-\rho_{\rm sec}^2}Q_4[(\rho_{\rm th}^{\rm comb})^2-2\rho_{\rm sec}^2]$, with $Q_4$ the survival function of the $\chi^2$ distribution with four degrees of freedom.

Equation~\eqref{eq:pairbudget} fixes $\rho_{\rm th}^{\rm comb}$ as a function of $\rho_{\rm sec}$. By Eq.~\eqref{eq:horizons}, our search then detects a pair down to the unlensed SNR
\be
\rho_{\rm UL,min} = \max\!\left[\frac{\rho_{\rm th}^{\rm comb}(\rho_{\rm sec})}{\sqrt{\mu_++|\mu_-|}},\,\frac{\rho_{\rm sec}}{\sqrt{|\mu_-|}}\right],
\label{eq:rhosec}
\ee
and we choose $\rho_{\rm sec}$ to maximize the volume $V\propto\langle\rho_{\rm UL,min}^{-3}\rangle$, averaged over the lens population with magnification bias. Two limits show what sets the result. The unlensed events alone must meet the budget, which requires approximately
\be
\rho_{\rm sec}^2 \gtrsim 2\ln\!\left[2N_{\rm delay}N_{\rm UL}(>\rho_{\rm th}^{\rm comb})/F\right],
\label{eq:rhosecfloor}
\ee
where $N_{\rm UL}(>\rho_{\rm th}^{\rm comb})$ is the number of unlensed events per run that pass the pair cut on their own. For unequal images, as for the point mass, the optimum lies close to this bound, and no $\rho_{\rm sec}<5.9$ meets the budget. For equal images ($a=1$), the two conditions of Eq.~\eqref{eq:horizons} coincide when $(\rho_{\rm th}^{\rm comb})^2=2\rho_{\rm sec}^2$, and the noise pairs then require $\rho_{\rm sec}^2\simeq\ln(N_{\rm pairs}/F)$, which lies above the bound of Eq.~\eqref{eq:rhosecfloor}. We use $\dd N/\dd\rho=3N_8\,8^3\rho^{-4}$, with $N_8=10^{2}$ detections per run above $\rho=8$. With the Morse phase imposed, the pair statistic is the maximum of $|z_L|^2$ in Eq.~\eqref{eq:recombine} over a real amplitude ratio of either sign, which has three instead of four degrees of freedom. Since $\sim\!10^{19}$ trials cannot be simulated, we compute these tail probabilities semi-analytically and check them against Monte Carlo draws. We recalibrate the targeted search on the same population, which gives $\rho_{\rm th}^{\rm sub}=6.17$ for a search after the detection and $6.28$ for one on both sides, as needed near a fold, where the order of the two images is unknown.

In the simulations of Sec.~\ref{sec:faint} and Appendix~\ref{app:pointmass}, we sample the sky position, polarization, inclination and arrival time of each pair, and compute the response of the LIGO Hanford detector at the arrival time of each image, including the rotation of the Earth over $t_d$. Compared with a fixed antenna pattern, this changes the volume ratios by less than $1\%$. For a fold the delay is short, so the response barely changes between the two images, and imposing the Morse phase loses nothing. For the overlapping images of Sec.~\ref{sec:sens}, the antenna pattern is common to the two images and to the unlensed template, so it cancels exactly in the volume ratios.

For a network of two detectors, each image has four degrees of freedom and the pair eight. We charge $N_{\rm sky}=10^{2}$ sky trials to the single-image search and to our pair search, and $N_{\rm loc}=10$ to the targeted search, which inherits the localization of its detection. The thresholds then rise by up to $\simeq1$, and the volume ratios change by at most $13\%$. For the point mass, the volume of our search changes from $2.17$ to $2.41$ times $V_{\rm sup}$, and from $0.97$ to $1.10$ times $V_{\rm one}$ in the analytic estimate from the horizons of Eq.~\eqref{eq:horizons}, because the brighter image of the targeted search must clear a higher network threshold.

\section{Fold caustics and the Chang--Refsdal lens}
\label{app:caustics}

Here we give the fold population of Sec.~\ref{sec:faintgain} and Fig.~\ref{fig:strategies}. Near a fold, at a distance $\delta$ of the source inside the caustic, the two merging images have magnifications $\mu_\pm\simeq\mu_f\,\delta^{-1/2}\pm\mu_0$, where $\mu_f$ and $\mu_0$ depend on the lens, so that both images have $\mu\propto\delta^{-1/2}$, the amplitude ratio has $1-a\propto\delta^{1/2}$, and their delay is $t_d\propto\delta^{3/2}$~\cite{Schneider:1992bmb}. For a source density uniform in $\delta$, a Euclidean population carries a magnification bias $\mu^{3/2}\propto\delta^{-3/4}$, so that in $u=1-a$ the population is $p(u)\propto u^{-1/2}$ for $0\le u\le u_{\max}$. Figure~\ref{fig:strategies} uses $u_{\max}=0.1$, i.e., $0.9\le a\le1$. Since $t_d\propto u^3$, a window $W$ selects $u_{\max}\propto W^{1/3}$, with a lens-dependent normalization.

We compute the volumes with a Monte Carlo simulation: on a grid of SNRs, we draw the complex SNRs of the two images, including the antenna pattern (Appendix~\ref{app:pairbackground}) and Gaussian noise, and apply the cuts of each strategy. The fraction of pairs passing is the detection efficiency $\varepsilon(\rho_\mathrm{UL})$, and $V\propto\int\varepsilon\,\rho_\mathrm{UL}^{-4}\,\dd\rho_\mathrm{UL}$. For two equally bright images ($a=1$), the volumes of Fig.~\ref{fig:strategies} change by at most $3\%$. For the wider population with $u_{\max}=0.3$, i.e., $0.7\le a\le1$, both our search and the targeted search gain more over requiring both images to be super-threshold, and our search has $1.2$--$1.7$ times the volume of the targeted search. For the targeted search we use $\rho_{\rm th}^{\rm sub}$ for a search on both sides of the detection (Appendix~\ref{app:pairbackground}). The targeted search gains near a fold only because of noise, which lets one of two equally bright images clear $\rho_{\rm th}$ while the other needs only $\rho_{\rm th}^{\rm sub}$; the sharp horizons of Eq.~\eqref{eq:horizons} miss this and give $V_{\rm one}=V_{\rm sup}$ at $a=1$, so we quote the simulation.

Fold pairs arise, e.g., for a compact lens embedded in the convergence and shear of a galaxy, i.e., the Chang--Refsdal lens of Sec.~\ref{sec:changrefsdal}. We sample its source plane uniformly in area and take the two brightest images. Inside the caustic, where folds dominate, the mean amplitude ratio is $\langle a\rangle=0.82$--$0.94$ across seven $(\kappa,\gamma)$ spanning both the macro-minimum and the macro-saddle, so these pairs are in the regime of Fig.~\ref{fig:strategies}. Outside the caustic the pairs are far more asymmetric, $\langle a\rangle=0.18$--$0.31$, and our search then has the same volume as the targeted search, as for the point mass.

\bibliography{references}

\end{document}